\documentclass[showpacs,preprintnumbers,amsmath,amssymb,prl,aps,twocolumn,
superscriptaddress, reprint,citeautoscript,fixfloat]{revtex4-1}
\usepackage{braket}
\usepackage[colorlinks=true, linkcolor=teal, citecolor=blue,
urlcolor=cyan]{hyperref}
\usepackage{xcolor,graphicx}
\usepackage{dcolumn}
\usepackage{amsmath}
\usepackage{amsfonts}
\usepackage{amssymb}
\usepackage{textcomp}
\usepackage{caption}
\usepackage[labelformat=empty]{subfig}
\usepackage{mathpazo}
\usepackage{eulervm}

\begin{document}

\title{New superexchange paths due to breathing-enhanced hopping   in
  corner-sharing cuprates}

\author{Nikolay A.~Bogdanov} %
\affiliation{Max Planck Institute for Solid State Research, Heisenbergstra{\ss}e
  1, 70569 Stuttgart, Germany}

\author{Giovanni Li~Manni} %
\affiliation{Max Planck Institute for Solid State Research, Heisenbergstra{\ss}e
  1, 70569 Stuttgart, Germany}

\author{Sandeep~Sharma} %
\affiliation{Max Planck Institute for Solid State Research, Heisenbergstra{\ss}e
  1, 70569 Stuttgart, Germany}
\affiliation{Department of Chemistry and Biochemistry, University of Colorado
  Boulder, Boulder, Colorado 80302, USA}

\author{Olle Gunnarsson} %
\affiliation{Max Planck Institute for Solid State Research, Heisenbergstra{\ss}e
  1, 70569 Stuttgart, Germany}

\author{Ali Alavi} %
\affiliation{Max Planck Institute for Solid State Research, Heisenbergstra{\ss}e
  1, 70569 Stuttgart, Germany} %
\affiliation{University Chemical Laboratory, Lensfield Road, Cambridge, CB2 1DEW,
  U.K.}

\date{\today}

\begin{abstract}
  We present {\it ab initio} calculations of the superexchange antiferromagnetic
  spin coupling $J$ for two cuprates, Sr$_2$CuO$_3$ and La$_2$CuO$_4$.
  Good agreement with experimental estimates is obtained.
  We find that $J$ increases substantially as the distance between Cu and apical
  O is increased.
  There is an important synergetic effect of the Coulomb interaction, expanding
  the Cu $3d$ orbital when an electron hops into this orbital, and the O-Cu
  hopping, being increased by this orbital expansion (breathing).
  This is a new ingredient in superexchange models.
  In a model with a fixed basis, breathing effects can be described as a mixing
  of $3d$ and $4d$ orbitals or as a single $3d \to 4d$ excitation.
\end{abstract}

\maketitle
Cuprates with corner-sharing CuO$_4$ plaquettes have received much attention due
to discoveries of high-temperature superconductivity \cite{Bednorz86} and exotic
states where spin and charge \cite{spin_charge_97} or spin and orbital
\cite{Sr213_RIXS} degrees of freedom are separated.
In these systems spins are coupled antiferromagnetically (AF), and it is
believed that spin fluctuations play an important role.
The strength of the AF coupling can be characterized by the nearest-neighbor
(NN) coupling $J$, which enters in, e.g., the Heisenberg model for undoped
systems and the $t$--$J$ model for doped systems \cite{ZhangRice_1988}.
Experimentally, the magnitude of $J$ varies strongly between different cuprates,
typically in the range 0.12-0.25~eV
\cite{Sr213_RIXS,La214_INS,Sr213_INS,La214_RIXS,J_apical_O_17}.%
This variation has been assigned to the dimensionality of the Cu-O network
\cite{Mizuno} and to the distance $d_{\mathrm{CuO}}$ between Cu and apical O
\cite{Mizuno,Weber}.
Our {\it ab initio} calculations show that an increase of $d_{\mathrm{CuO}}$
leads to substantially larger $J$.
Recent experimental work indeed finds such an increase \cite{Keimer}.

The AF coupling in these compounds occurs via superexchange
\cite{Kramers,Anderson}.
This involves the (virtual) electron hopping between the Cu $3d$ and O $2p$
orbitals.
Antiparallel spins on neighboring Cu atoms allow for more hopping possibilities
than parallel spins, leading to an AF coupling \cite{Khomskii_book2,SupMat}.
While the superexchange is well understood on the model level, the {\it ab
  initio} calculation of $J$ is a major problem.
For instance, calculations in a minimum though physically plausible basis set
underestimate $J$ by almost an order of magnitude.
This is therefore a long-standing problem in the \textit{ab initio} community
\cite{noci_J, ddci_La214, caspt2_ddci_J_de_Graaf_2001, Fink_J}.

We use wavefunction-based methods \cite{Helgaker_book} relying on full
configuration interaction quantum Monte Carlo (FCIQMC) \cite{FCIQMC_Booth_2009,
  Cleland_2011} and density-matrix renormalization group (DMRG)
\cite{DMRG_Sharma_2012, DMRGSCF_Sun_2017} as underlying solvers and restrict the
calculations to the NN $J$.
An exact calculation within our model would involve correlating $\approx 100$
electrons among $\approx 300$ orbitals, leading to a eigenvalue problem in a
Hilbert space of $10^{115}$ determinants.
Since problems on such a scale are out of reach, we use the method of
complete-active-space-self-consistent-field (CASSCF) together with
multi-reference perturbation theories to systematically approximate the
correlation energy \cite{CASSCF_Roos}.
In the CASSCF($n$,$m$) approach, a subset $n$ of the electrons (the active
electrons) are fully correlated among an active set of $m$ orbitals, leading to
a highly multi-configurational (CAS) reference wavefunction.
The choice of the active space will be discussed shortly, but let us note that
although this is still an exponential-scaling problem, it is manageable with the
aforementioned techniques as long as $n$ and $m$ are not too large.
In the SCF step, all orbitals are self-consistently optimised in the field of
the multi-determinant wavefunction (WF), to yield the variational minimum.
The CAS WF is then augmented using a number of second-order techniques,
including n-electron perturbation theory (NEVPT2) \cite{NEVPT2_method,icMRLCC},
multireference linearized coupled cluster (MR-LCC2) \cite{MRLCC_method,icMRLCC},
or multireference configuration interaction with single and double excitations
(MR-CISD) \cite{Helgaker_book}; these methods capture the remaining (weak)
correlation involving electrons and orbitals outside of the active space.
We use these different second order methods as a gauge of their reliability.
As the active space is enlarged, the corresponding second-order corrections
diminishes.
The key question that arises is: what is the ``minimal'' active space necessary
to obtain a qualitatively correct reference wavefunction, sufficient to compute
$J$ reliably?
We find that the necessary active space needs to be far larger than previously
imagined, including relatively high energy Cu $4d$ and O $3p$ orbitals.
Exclusion of these states from the active spaces leads a dramatic
under-estimation of $J$.

We analyze the reason for the strong dependence of $J$ on the active space and,
in particular, the importance of $4d$ orbitals.
As mentioned above, the superexchange mechanism depends on O-Cu hopping.
The Coulomb energy cost $U_{\rm eff}$ of this hopping is strongly reduced by an
expansion of the Cu $3d$ orbitals, referred to as breathing \cite{U}, when an
electron hops into a Cu $3d$ orbital.
This breathing effect at the same time increases the Cu-O effective hopping
integral $t_{\rm eff}$ \cite{hopping}.
In a similar way the O $2p$ orbital breathes as the O occupancy is changed.
In the superexchange mechanism, $J$ depends on both $U_{\rm eff}$ and $t_{\rm
  eff}$ \cite{SupMat}
and the breathing effects therefore strongly influence $J$.
In a Hilbert space constructed with orthogonal orbitals, the breathing effect is
described by the formation of linear combinations of, e.g., $3d$ and $4d$
orbitals, showing up as an increased occupancy of the $4d$ orbitals.

The breathing effects involve a single $3d\to 4d$ excitation, leading to an
{\it expansion} of the charge density when an electron is added to the $d$
shell.
There are also important double $3d\to 4d$ excitations, which provide radial
(in-out) correlations \cite{Helgaker_book}.
For a fixed number of $d$ electrons, these correlations lead to a {\it
  contraction} of the charge density, at least if the basis has sufficient
flexibility to satisfy the virial theorem.
Correlation and breathing compete, making the simultaneous description
complicated.
Both effects lead to the occupancy of $4d$ orbitals, but are otherwise very
different.

To study the electronic structure of cuprates we employ the embedded cluster
model.
With this approach accurate high-level calculation is performed for a small
representative unit of the solid, while its environment is treated in a more
approximate manner \cite{de_Graaf_Broer_book}.
We use clusters that include two CuO$_4$ (CuO$_6$) units, two (ten) neighboring
Cu$^{2+}$ ions and all adjacent Sr$^{2+}$ (La$^{3+}$) ions, in total
[Cu$_4$O$_7$Sr$_{16}$] and [Cu$_{12}$O$_{11}$La$_{16}$] for Sr$_2$CuO$_3$ and
La$_2$CuO$_4$ respectively.
The rest of the solid is modeled by an array of point charges fitted to
reproduce the Madelung potential in the cluster region \cite{Ewald_method,
  Ewald_soft, madpot_soft}.
We employed the crystal structures as reported in Refs.~\cite{str_Sr213} and
\cite{str_La214}.
The value of the NN superexchange parameter can be easily extracted by mapping
the energy spectrum of the two-magnetic-site cluster to two-site Heisenberg
model.
To make this mapping straightforward, the peripheral Cu ions are represented by
total-ion potentials with no associated electrons, such that $J$ can be
extracted as the energy difference of lowest triplet and singles states
\cite{de_Graaf_Broer_book}.
We use all electron \mbox{cc-pVDZ} and \mbox{cc-pVTZ}o basis sets for central Cu
and O ions \cite{basis_O, basis_Cu}, large-core effective potentials for other
species \cite{basis_Sr, basis_La1,basis_La2} and utilize several quantum
chemistry computational packages \cite{NECI,molcas8,pyscf,molpro12,columbus},
see supplementary material for more details \cite{SupMat}.

We first perform CASSCF calculation with two singly occupied Cu 3$d_{x^2-y^2}$
orbitals in the active-space, CASSCF(2,2)
similar to the one-band Hubbard model.
Such minimal active-space calculation accounts for the unscreened Anderson
superexchange mechanism ($d^9-d^9$ and $d^{8}-d^{10}$ configurations) and gives
a qualitatively correct AF $J$ coupling.
The value of the $J$ obtained this way is, however, only $\approx\!\!20\%$ as
compare to the experimental data \cite{Sr213_RIXS, Sr213_INS, La214_RIXS,
  La214_INS}, see \autoref{tab:J_table}.
As it can be seen, the second order corrections nearly double $J$, but are
clearly an insufficient treatment.
Such uniform behavior of the different dynamical correlation methods suggests
that the reference CASSCF calculation is inadequate to qualitatively describe
the system, and that the active space has to be enlarged.
The only exception is the difference-dedicated configuration interaction (DDCI)
method that gives values of $J$ very close to experiment on top of CASSCF(2,2)
reference \cite{ddci_La214, DDCI_note}.
However, the DDCI is essentially a subspace of the MRCI-SD, and significant
differences of $J$ calculated by these two methods imply that the description of
electronic structure given by DDCI is far from being complete.

\begin{table}[t]
  \caption{
    Values of the superexchange parameters $J$ (in~meV) obtained with different
    methods, see the text.
  }
  \label{tab:J_table}
  \begin{ruledtabular}
    \begin{tabular}{lcc|c}
            &\multicolumn{2}{c|}{Sr$_2$CuO$_3$}
                                 &La$_2$CuO$_4$\\
                &cc-pVDZ &cc-pVTZ &cc-pVDZ\\
\hline
&&&\\[-0.15cm]
CASSCF(2,2)   &37      &36      &34\\
~~~~\textnormal{+MR-LCC2}
                &\textnormal{57}
                         &\textnormal{68}
                                  &\textnormal{52}\\
~~~~+MR-CISD    &60      &52      &50\\
~~~~+NEVPT2     &67      &78     &59\\
~~~~+DDCI       &\,246\footnotemark[1]
                         &274     &\,150\footnotemark[1]\\
\hline
&&&\\[-0.15cm]
CASSCF(4,3)     &39      &38      &35\\
~~~~+MR-CISD    &105      &93      &\\
~~~~+NEVPT2     &112      &135    &61\\
\hline
&&&\\[-0.15cm]
CASSCF(8,10)  &70      &69     &50\\
~~~~\textnormal{+MR-LCC2}
                &\textnormal{151}
                         &\textnormal{148}
                                 &\textnormal{112}\\
~~~~+MR-CISD    &153     &--     &--\\
~~~~+NEVPT2     &145     &143    &107\\
~~~~+CASPT2     &\;260\footnotemark[1]
                         &205    &\;139\footnotemark[1]\\
\hline
&&&\\[-0.15cm]
CASSCF(24,26) &125     &116  &90\\
~~~~\textbf{+MR-LCC2}
                &\textbf{252}
                         &\textbf{256}
                               &\textbf{145}\\
~~~~+NEVPT2     &253     &262     &148\\
\hline
&&&\\[-0.15cm]
experiment  &\multicolumn{2}{c|}{249\cite{Sr213_RIXS}, 241\cite{Sr213_INS}}
                              &120\cite{La214_RIXS}, 138\cite{La214_INS}\\
    \end{tabular}
    \footnotetext[1]{Ref.\,\onlinecite{caspt2_ddci_J_de_Graaf_2001},
      calculations performed with different clusters and basis sets}%
  \end{ruledtabular}
\end{table}

Because an electron hopping from the bridging O $\sigma$-bonding 2$p_y$ orbital
to the Cu 3$d_{x^2-y^2}$ plays a crucial role in the superexchange (see, e.g.,
Ref.\,\onlinecite{Khomskii_book2}), this orbital is an obvious candidate to add
into the active space.
Such CASSCF(4,3) calculation roughly corresponds to an unscreened 3-band Hubbard
model.
However, the obtained magnetic couplings turn out to be less than 1 meV higher
compared to CASSCF(2,2).
The reason is that, despite the inclusion of important ligand-hole determinants
($d^{9}$-$p^5$-$d^{10}$ and $d^{10}$-$p^4$-$d^{10}$), their energy is too high
to be effective, as the orbital optimization is primarily driven by the dominant
$d^{9}$-$p^6$-$d^{9}$ configuration \cite{noci_J, caspt2_ddci_J_de_Graaf_2001,
  J_Malrieu_review_2013}.
When we include the effect of further excited determinants at 2nd-order level on
top of the CASSCF(4,3) WF, $J$ becomes significantly larger, 105 eV using
MR-CISD for the Sr$_2$CuO$_3$ compound.
It is still more than two times smaller than the experimental value, indicating
that important details are still missing.

To give the WF flexibility for accounting the orbital relaxation in
$d^8$-$p^6$-$d^{10}$ and $d^{9}$-$p^5$-$d^{10}$ determinants, one can add a set
of orbitals previously kept empty (Cu 4$d$ and O 3$p$) to the active space
\cite{noci_J, caspt2_ddci_J_de_Graaf_2001}.
Having additional $d$ orbitals in the active space has been shown to be
necessary to describe multiplet splittings for the late transition metals of the
first row, see, e.g., Refs \onlinecite{Roos_1992, Pierloot_2011}.
Due to variationality of orbital optimization within the CASSCF procedure, the
active orbitals are allowed to change, and a balanced choice of active space is
required to ensure convergence.
Balanced active space can be constructed with Cu 3$d$ and 4$d$ orbitals of $e_g$
character plus the bridging oxygen 2$p_y$ and 3$p_y$ orbitals
\cite{caspt2_ddci_J_de_Graaf_2001}.
Results for the CASSCF(8,10) calculations are shown in the third block of the
\autoref{tab:J_table}.
The extention of the active space leads to a systematic differential effect, $J$
significantly increases at all levels of theory.
Results close to experiment were reported using this active space together with
a different formulation of the perturbation theory
\cite{caspt2_ddci_J_de_Graaf_2001}.
But our calculations give 60 and 80\,\% of the experimental values for the Sr
and La cuprates respectively.
\begin{figure}[t]
  \includegraphics[width=.96\linewidth]{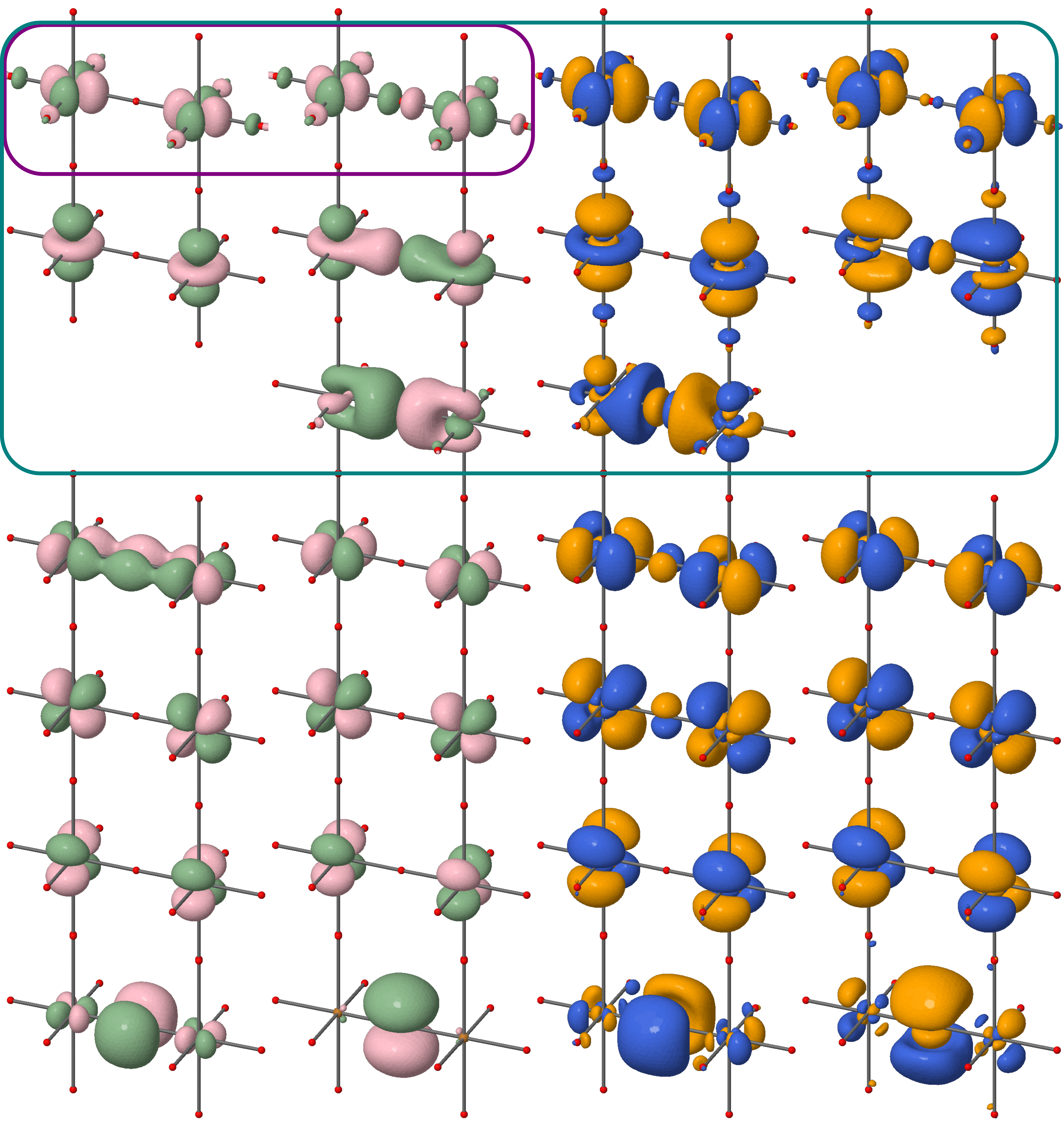}
  \caption{ %
    Orbitals optimized in the CASSCF(24,26) calculation for the La$_2$CuO$_4$
    compound.
    (2,2) and (8,10) subsets are indicated by smaller rectangles.
  }
  \label{fig:orbitals}
\end{figure}
To get the balanced description of all relevant effects we add all copper 3$d$
and 4$d$ together with the bridging oxygen 2$p$ and 3$p$ orbitals, resulting in
CASSCF(24,26).
This active space yields diagonalization problem in the space of
$\approx\!10^{14}$ Slater determinants and is not feasible with conventional
diagonalization methods, therefore we have to proceed with DMRG and FCIQMC as
approximate solvers \cite{FCIQMC-CASSCF, DMRGSCF_Sun_2017, SupMat}.
With the additional many-body contribution from the Cu $t_{2g}$ and
$\pi$-bonding O orbitals taken into account in the large CASSCF we find further
stabilization of the singlet compare to the triplet.
Second order correction on top of the CASSCF(24,26) reference finally brings $J$
close to the experimental values, see the last block in \autoref{tab:J_table}.
Orbitals as optimized in the variational calculation are shown in
\autoref{fig:orbitals}.
One can notice that both 3$d$ and 4$d$ orbitals have significant amplitudes at
the bridging oxygen $p$ orbitals.
%
%

In the standard theory \cite{Kramers,Anderson} of superexchange, a model of
Cu$_2$O is treated, with one nondegenerate orbital on each atom.
As discussed above, including only these orbitals in CASSCF(4,3) underestimates
$J$ by almost one order of magnitude.
We now discuss why it is necessary to consider the large active space.

In calculations for the singlet and triplet states, the configuration
$d^9$-$p^6$-$d^9$ dominates, and correlation effects in this configuration are
very important for the total energy.
However, the contributions for the singlet and triplet states are similar and
therefore not very important for $J$.
In superexchange theory, hopping of electrons between Cu and O plays a crucial
role, involving, e.g., $d^{8}-d^{10}$ determinants in addition to the nominal
$d^9-d^9$.
There are more configurations of this type available for the singlet than the
triplet state.
Although these configurations have rather small weights, they are crucial for
$J$.
These configurations are indicated 
in the supplementary material \cite{SupMat}.

The on-site Coulomb integral between two $3d$ electrons is very large
($\approx\!28$\,eV \cite{footnot_atomicU}), leading to drastically suppressed
charge fluctuations in the simplest model.
This is the reason the CASSCF(2,2) and CASSCF(4,3) give a very small $J$.
However, by increasing the active space size, this energy cost can be strongly
reduced.
Crucial effects are the change of the effective radial extent of the $3d$
orbital (breathing) and rearrangements of the non-$3d$ charge density as the
number of $3d$ electrons varies (screening) \cite{U}, which are captured in the
CASSCF(24,26) calculation with second-order correction.

In order to disentangle these different effects we performed a series of simpler
constrained calculations \cite{GASSCF, SupMat}.
We put all hopping integrals from $d$ ($3d$ or $4d$) basis functions on the Cu
atoms equal to zero.
We can then prescribe the total occupancy of $d$ orbitals on each Cu atoms.
We performed two calculations, one with the configurations $d^9-d^9$ and one with
$d^8-d^{10}$.
In both cases the system is allowed to fully relax, except that hopping to or
from $d$ orbitals is suppressed.
We then obtain that the energy for the $d^8-d^{10}$ state is about 12~eV higher
than for $d^9-d^9$.
This means that $U\!\approx\!28$~eV has been reduced to
$U_{\mathrm{eff}}\!\approx\!12$~eV.
According to other estimates, $U_{\rm eff}$ is reduced even further
($\approx\!8$~eV \cite{Cu}).


\begin{figure}[t]
  \includegraphics[width=.99\linewidth]{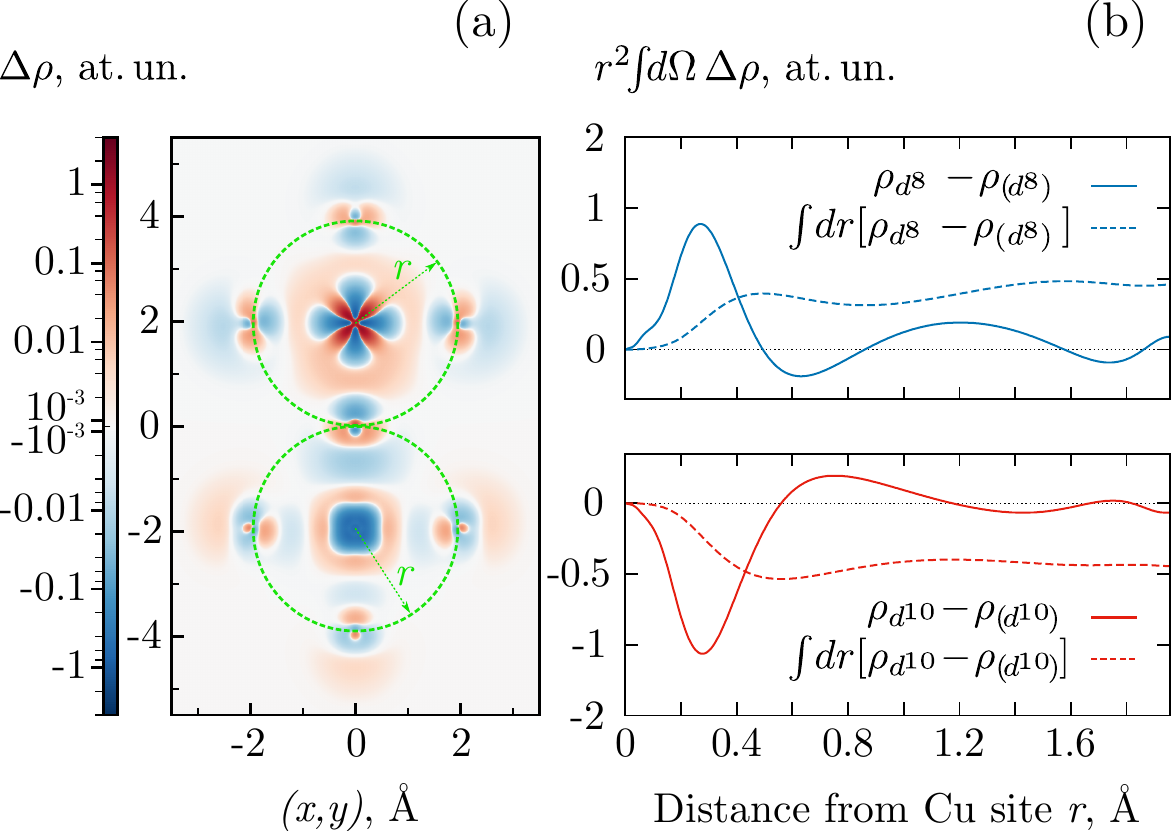}
  \caption{Electron density difference due to orbital relaxation in the
    $d^8-d^{10}$ configuration.
    (a) 
    In the plane of CuO$_4$ plaquettes;
    (b) 
    integrated over a sphere centered on one of the Cu
    atoms (full curves) as a function of the radius as  shown in (a).
    The results of an additional radial integration are shown by the dashed
    curves as a function of the upper integration limit.
  }
  \label{fig:chargediff}
\end{figure}

\autoref{fig:chargediff} shows charge differences due to breathing and screening
for the $d^8-d^{10}$ calculation, discussed above.
A calculation was first performed for $d^9-d^9$, then a $d$ electron was moved
from one Cu atom to the other keeping all orbitals unchanged.
The corresponding densities are denoted $\rho_{(d^8)}$ and $\rho_{(d^{10})}$.
This $d^8-d^{10}$ state was then allowed to relax self-consistently, giving the
densities $\rho_{d^8}$ and $\rho_{d^{10}}$.
The full red curve shows the change in charge density
$\rho_{d^{10}}-\rho_{(d^{10})}$, illustrating how charge is moved from the inner
part of Cu to the outer part (breathing).
The red dashed curve shows a radial integral of the charge density difference.
It shows that more charge is removed from the inner part than added to the outer
part.
Since the number of $d$-electrons is the same in the two calculations, non-$d$
charge has been moved away from the Cu atom with the $d^{10}$ configurations as
a response to the addition of one $d$ electrons (screening).
Adding a $d$ electron to a Cu atom then only leads to an increase of the net
charge by about half an electron, due to screening, which substantially reduces
the energy cost.

As in can be seen in \autoref{tab:J_table}, magnetic coupling in Sr$_2$CuO$_3$
is nearly two times larger than in La$_2$CuO$_4$.
In both cases the computation of $J$ is done using only two magnetic centers,
therefore this difference should not be attributed to dimensionality of two
materials.
The other structural difference is the presence of apical oxygen ions in the
La$_2$CuO$_4$ that changes the local multiplet splittings, mainly the position
of 3$d_{z^2}$ levels \cite{Sr213_RIXS, CuO2_rixsdd_11, Hozoi_SciRep, Huang_PRB}.
Relative energy of 3$d_{z^2}$ orbital is believed to be connected to the shape
of the Fermi surface and the value of the critical temperature in doped cuprates
\cite{CuO2_z2_eskes91, CuO2_z2Tc_ohta91, CuO2_z2_feiner96, CuO2_QPs_hozoi08,
  CuO2_QPs_hozoi08}.

\begin{figure}[t]
  \includegraphics[width=.96\linewidth]{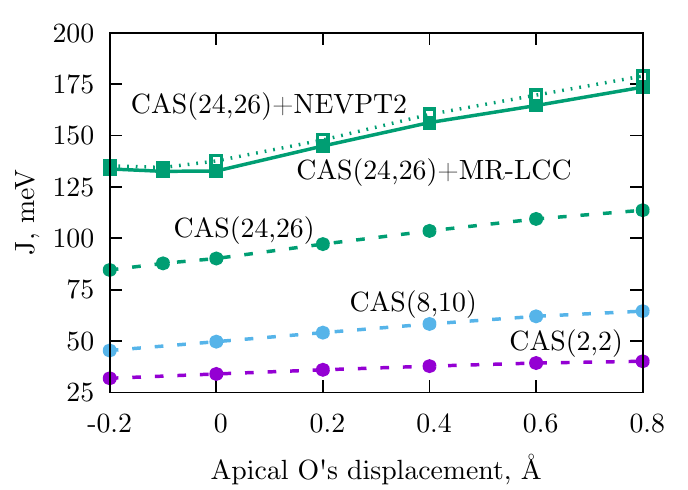}
  \caption{ %
    Dependence of $J$ on the distance to apical oxygens in La$_2$CuO$_4$.
  }
  \label{fig:La214_JvsDist}
\end{figure}

There are experimental evidences that $J$ also changes depending on the local
geometry \cite{J_apical_O_17}.
However, because experimentally different compounds have to be used, local
distortions are accompanied by changes of Cu--O distances and type of adjacent
metal ions.
Therefore it is instructive to investigate the dependence of $J$ on the distance
to apical oxygen ions in La$_2$CuO$_4$ compound with accurate computational
method.
We varied the apical O's positions within the cluster keeping the electrostatic
potential untouched and compute magnetic couplings using the procedure described
above.
The results of these calculations are presented in \autoref{fig:La214_JvsDist}.
It can be seen that with increase of the distance to apical oxygen the NN $J$
grows.
It shows that the growth is faster with more electron correlation is taken into
account.
One obvious effect that leads to increase of $J$ is the lowering of the
4$d_{z^2}$ orbital energy and corresponding enhancement the orbital breathing:
we observe 13\% growth of the occupation of 4$d_{z^2}$ orbitals upon
0.8\,\AA~displacement of apical oxygens at the CASSCF(24,26) level.
Recent experimental results for $J$ in thin films is in good agreement with our
calculations
\cite{Keimer}.

In this letter we presented state-of-the are \textit{ab initio} calculations of
the NN magnetic coupling in cuprate compounds.
We find that orbital breathing involving Cu 4$d$ orbitals is essential to
properly describe magnetism in these materials.
We find that a synergistic coupling between orbital breathing and enhanced
hopping lead to the observed $J$.
This mechanism also leads to a strong dependence of $J$ on the distance to
apical oxygen ions, where a
30\% increase of the coupling occurs under 30\% elongation of the Cu--O
distance.
More generally, breathing-enhanced hopping may be expected to play an important
role in generating longer range hopping and exchange interactions, beyond
nearest neighbors.
Significant long range hopping would provide a mechanism to generate frustration
on the square lattice, and in the doped cuprates may be relevant in the
mechanism leading to superconductivity.
These questions are left for a future study.


\bibliography{refs}

\begin{thebibliography}{63}
\bibitem [{foo()}]{footnot_atomicU}%
  \BibitemOpen
  \bibinfo {note} {Computed as $U\!=\!2E_{d^9}\!-\!E_{d^8}\!-\!E_{d^{10}}$ for
  Cu ion at CASCI level with 9, 8 and 10 electrons in 5 orbitals optimized for
  the $d^9$ state. Bare Coulomb integral $\left < 3d 3d \| 3d 3d \right >$ is
  29.8\,eV in these calculations.}\BibitemShut {Stop}%
\end{thebibliography}

\clearpage

\setcounter{figure}{0}
\setcounter{table}{0}
\setcounter{equation}{0}
\renewcommand{\thefigure}{S\arabic{figure}}
\renewcommand{\thetable}{S\Roman{table}}
\renewcommand{\theequation}{S\arabic{equation}}

\def\equationautorefname~#1\null{%
  Eq.~(#1)\null
}
\def\figureautorefname~#1\null{%
  Fig.~#1\null
}

\makeatletter
\DeclareRobustCommand{\element}[1]{\@element#1\@nil}
\def\@element#1#2\@nil{%
  #1%
  \if\relax#2\relax\else\MakeLowercase{#2}\fi}
\pdfstringdefDisableCommands{\let\element\@firstofone}
\makeatother

\section{Supplementary material}\label{SupMat}
\section{Details of calculations}\label{sec:details}

For small CASSCF calculations up to (8,10) active space were done with MOLCAS,
MOLPRO and PySCF programs \cite{molcas8, molpro12, pyscf}.
Results by different codes are fully consistent, differences of total energies
were not more than $10^{-6}$ Hartree.
All NEVPT2 and MR-LCC2 calculations were carried out with IC-MPS-PT and BLOCK
programs \cite{icMRLCC, DMRG_Sharma_2012}.
MRCI-SD calculations were done using COLUMBUS and MOLCAS driver \cite{columbus,
  molcas8}.
CASPT2 calculations were performed with MOLCAS 8 \cite{molcas8}.
DDCI calculations were done with MRCI module of MOLPRO \cite{molpro12}.
Large CASSCF(24,26) calculations were carried out with CASSCF module of MOLCAS
using NECI as a solver \cite{FCIQMC-CASSCF, molcas8, NECI} and independently
with PySCF using BLOCK as a solver \cite{pyscf, DMRG_Sharma_2012}.

Data shown in \autoref{fig:chargediff} was obtained in constrained calculations
using generalized active space SCF (GASSCF) method as implemented in MOLCAS
\cite{GASSCF, molcas8}.
We split atomic-like orbitals in three groups: all $d$ orbitals at the first
copper ion (15 in cc-pVDZ basis), all $d$ orbitals at the second copper ion
(15), and the rest.
Any orbital rotation between these groups are forbidden via supper-symmetry
constrain.
With GASSCF we specify two disconnected active spaces, e.g., (8,5) and (10,5)
for the first and the second Cu ion respectively.
This way we can fix occupation of $d$ orbitals at each site and perform all the
possible remaining optimizations.

Density plots were done with the Multiwfn program \cite{Multiwfn}.
Molecular orbitals were plotted with Jmol \cite{Jmol}.

\onecolumngrid

\begin{figure*}[h]
  \includegraphics[width=0.90\linewidth]{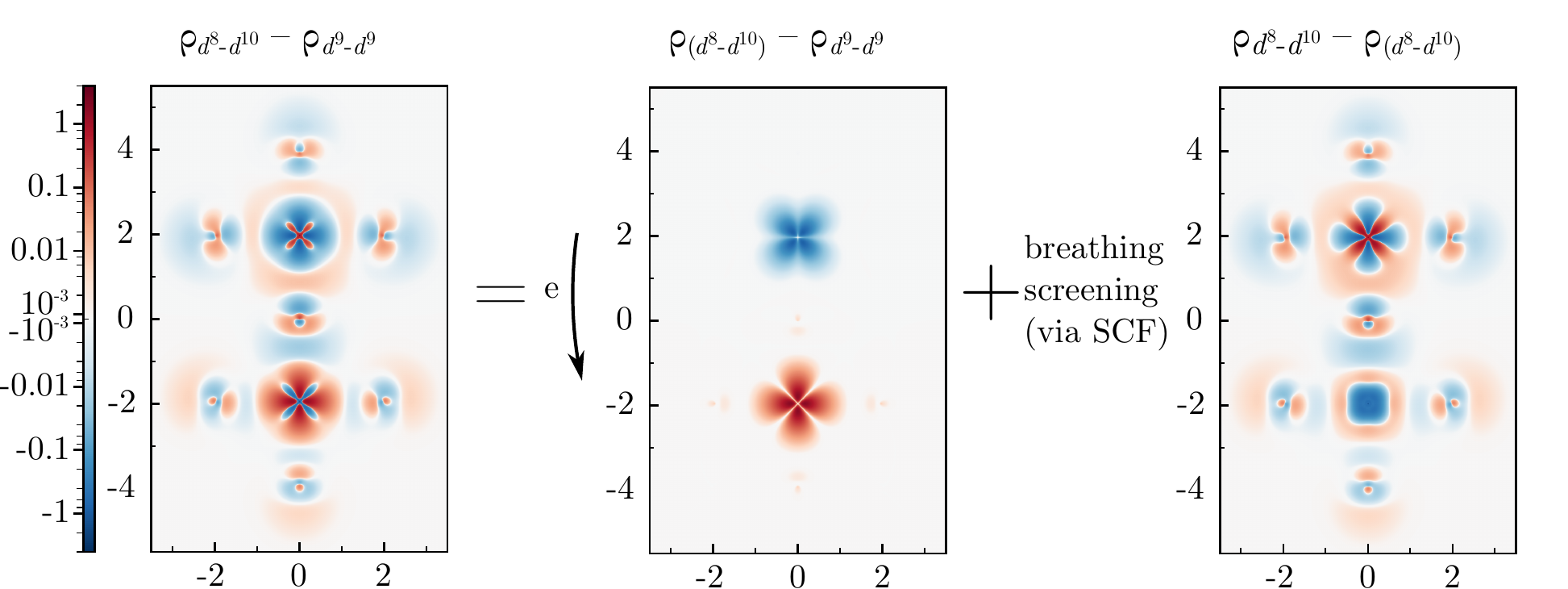}
  \caption{Electron density differences as obtained from constrained
    calculations for Sr$_2$CuO$_3$.
  }
  \label{fig:dDend3}
\end{figure*}
\twocolumngrid

\begin{figure}[b]
  \includegraphics[width=.82\linewidth]{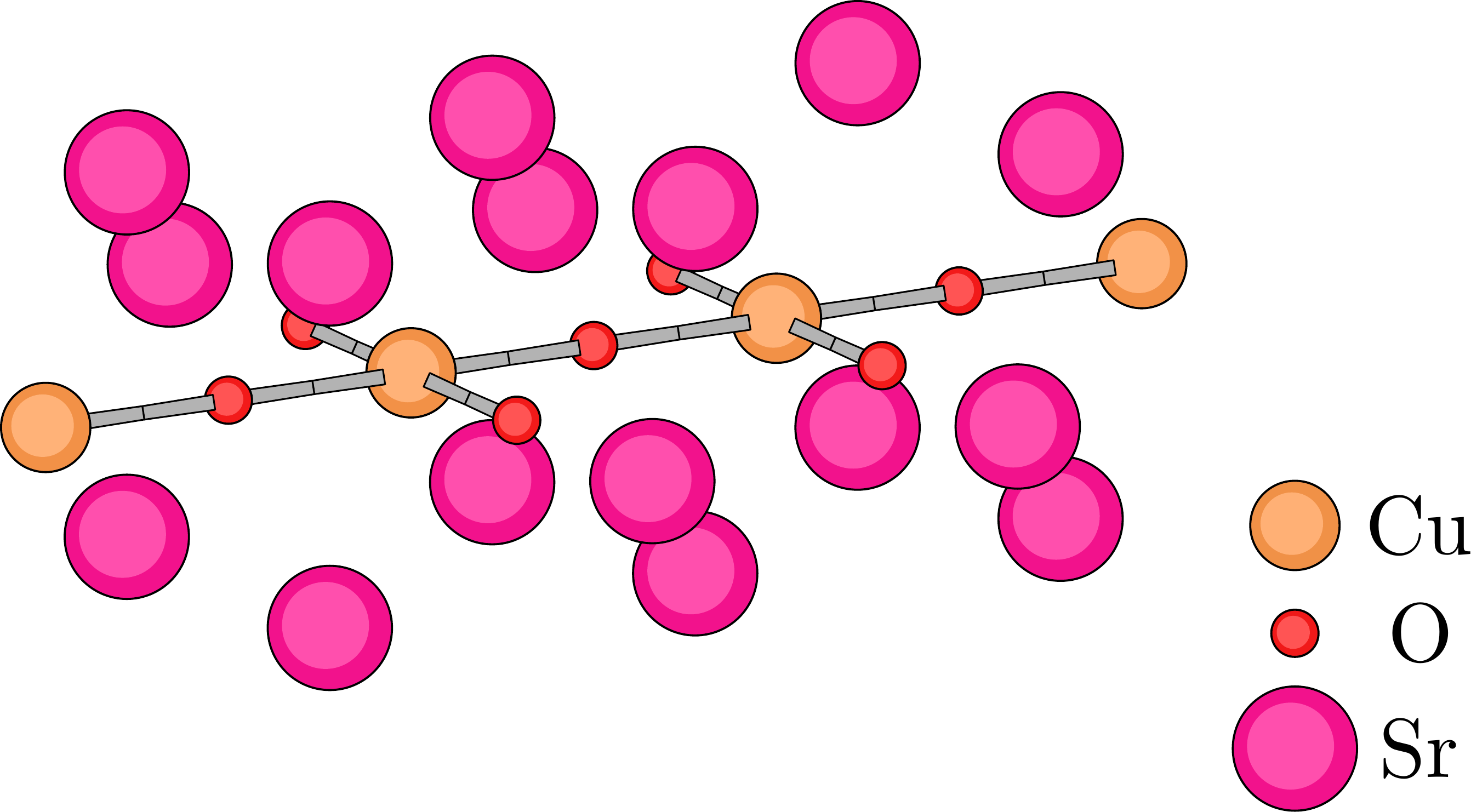}
  \caption{ Sketch of the [Cu$_4$O$_7$Sr$_{16}$] cluster used in calculations of
    the Sr$_2$CuO$_3$ compound.
  }
  \label{fig:Sr213_cluster}
\end{figure}

\begin{figure}[t]
  \includegraphics[width=.99\linewidth]{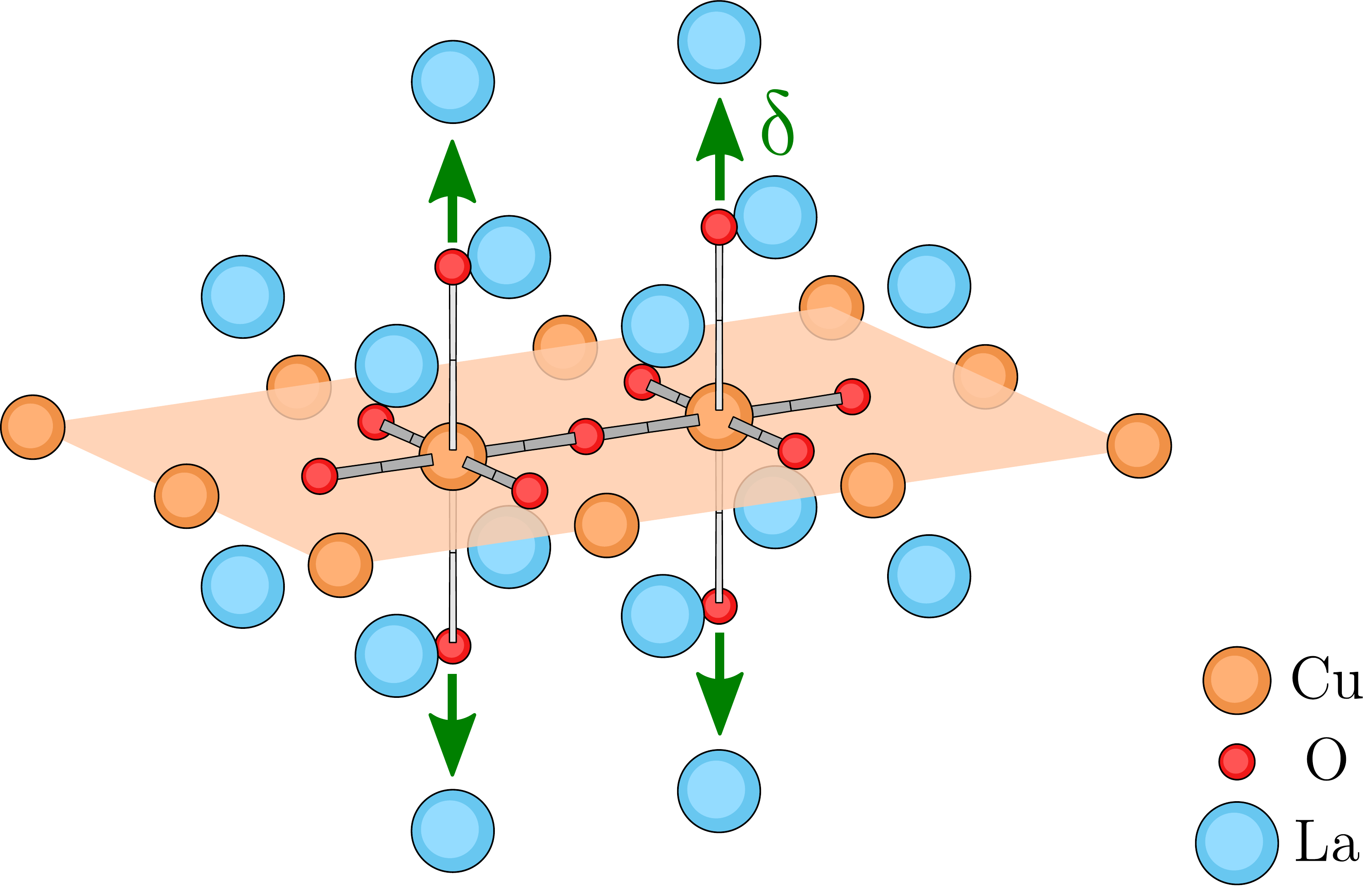}
  \caption{
    Sketch of the [Cu$_{12}$O$_{11}$La$_{16}$] cluster used in calculations of
    the La$_2$CuO$_4$ compound.
    Green arrows represent the apical oxygen displacements discussed in the main text.
  }
  \label{fig:La214_cluster}
\end{figure}

Clusters used in calculations are shown in \autoref{fig:Sr213_cluster} and
\autoref{fig:La214_cluster}, both belong to the $D_{2h}$ point group.
In calculations with MOLCAS, NECI, COLUMBUS, PySCF, and BLOCK were done within
the full point-group symmetry.
Irreducible-representation composition of active orbitals reads CAS(2,2):
[1$a_g$, 1$b_{2u}$]; CAS(4,3): [1$a_g$, 2$b_{2u}$]; CAS(8,10): [4$a_g$,
6$b_{2u}$]; CAS(24,26): [4$a_g$, 2$b_{1g}$, 2$b_{2g}$, 2$b_{3g}$, 2$a_u$,
4$b_{1u}$, 6$b_{2u}$, 4$b_{3u}$].
In our setting the Cu-O-Cu link is along the $y$ direction.
Due to technical limitations calculations with MOLPRO were done within the $C_1$
point group.

All NEVPT2, MR-LCC2, and MRCI calculations were done correlating all Cu $d$ and
O $p$ electrons, which is in total 60 and 84 electrons for Sr$_2$CuO$_3$ and
La$_2$CuO$_4$ respectively.

To illustrate better the constrained GASSCF calculations presented in the main
text we show in \autoref{fig:dDend3} electron density differences between three
states.
We start with fully optimized $d^9-d^9$ state, then obtain $(d^8-d^{10})$ state
by moving an electron from one Cu site to the other keeping the orbitals
untouched, and finally reach the $d^{8}-d^{10}$ state after orbital relaxation
that captures breathing and screening.
One should notice that in both latter states multiplet effects at the $d^{8}$
site are taken into account by including all determinants that arise by
distributing 8 electrons in five $d$ orbitals into the WF expansion.
This multiplet effects lead to $\approx\!1$\,eV reduction of the
$U_{\mathrm{eff}}$.

\section{Breathing in a \element{Cu}$_2$ model}\label{sec:cu2}
In this section we discuss breathing in a very simple Cu$_2$ model, with an
effecting hopping directly between two Cu atoms rather than via bridging O.
We show how the radial extent of the Cu 3$d$ orbital is effectively increased in
intermediate states with increased 3$d$ occupancy.
This has two important consequences.
First, the effective energy cost of increasing the occupancy of 3$d$ level is
reduced, since the electrons can avoid each other better\cite{U}.
Second, the hopping between the two sites is enhanced, as the Cu 3$d$ orbital
expands\cite{hopping}.

\begin{figure}[!t]
  \includegraphics[width=.325\linewidth]{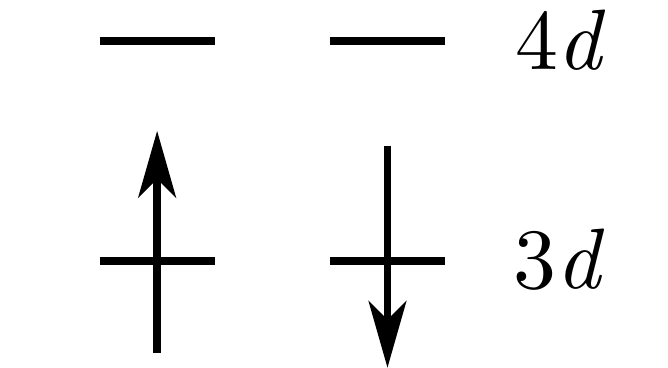}
\caption{Schematic representation of the Cu$_2$ dimer with 3$d$ and
4$d$ levels.\label{fig:dimer}}
\end{figure}

In the CASSCF calculations in the main text, a fixed orthogonal basis set is
used for all intermediate states.
Therefore the breathing effect of a 3$d$ orbital is then described as a mixing
of the 3$d$ and 4$d$ orbitals.
The system can then effectively expand or contract an effective 3$d$ orbital,
depending on the relative sign of the mixing.
To illustrate how this happens we consider a Cu$_2$ dimer, including just one
3$d$ and one 4$d$ level on each atom, as indicated in \autoref{fig:dimer}.
The levels have spin but no orbital degeneracy.
We use the Hamiltonian 
\begin{widetext}
\begin{eqnarray}\label{eq:2}
  H&&=\sum_{\sigma} \left[
    \sum_{i=1}^2 \sum_{j=1}^2 \varepsilon_j n_{ij\sigma} +
    \sum_{i=1}^2\sum_{j=1}^2 t_{ij} \left(
      c^{\dagger}_{1i\sigma}c^{\phantom \dagger}_{2j\sigma}+
      c^{\dagger}_{2i\sigma}c^{\phantom \dagger}_{1j\sigma}
    \right)
  \right] 
    +\sum_{i=1}^2 \left[
    U_{11}n_{i1\uparrow}n_{i1\downarrow}+U_{22}n_{i2\uparrow}n_{i2\downarrow}
    +U_{12}\sum_{\sigma \sigma'}n_{i1\sigma}n_{i2\sigma'}
  \right] \nonumber \\
  &&+\sum_{i=1}^2\sum_{\sigma} \Big(K_1n_{i1\sigma}+K_2n_{i2\sigma} \Big)
  \left(c^{\dagger}_{i1-\sigma}c^{\phantom \dagger}_{i2-\sigma}+
    c^{\dagger}_{i2-\sigma}c^{\phantom \dagger}_{i1-\sigma}\right).
\end{eqnarray}
\end{widetext}
Here the first index on $c_{ij\sigma}$ refers to the site and the second labels
the orbital, i.e., $j=1 (2)$ refers to a 3$d$ (4$d$) orbital.
The hopping between the Cu atoms is described by $t_{ij}$.
We also include the direct on-site Coulomb integrals $U_{11}$, $U_{12}$ and
$U_{22}$, describing $3d-3d$, $3d-4d$ and $4d-4d$ interaction, respectively.
$K_i$ refers to a Coulomb integral with three equal orbitals and the fourth
different:
\begin{align}\label{eq:3}
  &&K_i=e^2\int d^3 r\int d^3 r'
  {\phi_i({\bf r})^2 \phi_1({\bf r'})\phi_2({\bf r'}) \over |{\bf r}-{\bf r}'|}.
\end{align}
These integrals are crucial for the breathing effect.
If, e.g., the 3$d$ orbital on an atom is doubly occupied, the last term in
\autoref{eq:2} can excite a single electron from the 3$d$ orbital to the 4$d$
orbital.
For radial (in-out) correlation it is important to include terms where two 3$d$
electrons are excited to the 4$d$ level.
Such terms are neglected here.

For simplicity, we here put $t_{12}=t_{21}=\sqrt{t_{11}t_{22}}$,
$U_{12}=\sqrt{U_{11}U_{22}}$ and $K_1/K_2=\sqrt{U_{11}/U_{22}}$.
We have used \mbox{$\varepsilon_2-\varepsilon_1=24$~eV}, $U_{11}=13$~eV,
$U_{22}=10$~eV, $K_1=-8$~eV, $t_{11}=-0.5$~eV and $t_{22}=-0.8$~eV.

\subsubsection{Free atom}\label{sec:cu2.atom}
To study the breathing effect, we first consider a free atom, setting
$t_{11}=t_{12}=t_{21}=t_{22}=0$ in \autoref{eq:2}.
We put two electrons on one site and write down the wave-function
\begin{eqnarray}\label{eq:4}
  |\Phi\rangle\,&&=\bigg[a_{11}c^{\dagger}_{11 \uparrow}c^{\dagger}_{11 \downarrow}
  +a_{12}\left(c^{\dagger}_{11 \uparrow}c^{\dagger}_{12 \downarrow}
  +c^{\dagger}_{12 \uparrow}c^{\dagger}_{11 \downarrow}\right) \nonumber \\
  &&+\,a_{22}c^{\dagger}_{12 \uparrow}c^{\dagger}_{12 \downarrow}\bigg] |{\rm vac}\rangle  \\
  &&=b^2\tilde c^{\dagger}_{11\uparrow}\tilde c^{\dagger}_{11\downarrow}|{\rm vac}\rangle
  +\left(a_{22}-{a_{12}^2\over a_{11}}\right)
  c^{\dagger}_{12 \uparrow}c^{\dagger}_{12 \downarrow} |{\rm vac}\rangle \nonumber
\end{eqnarray}
Apart from the last term, we have replaced the 3$d$ orbital by an expanded
orbital described by $\tilde c_{11\sigma}^{\dagger}$
\begin{equation}\label{eq:4a}
  \tilde c_{11\sigma}^{\dagger}=
  {1\over b}\left(
    \sqrt{a_{11}}c^{\dagger}_{11\sigma}+{a_{12}\over \sqrt{a_{11}}}c^{\dagger}_{12\sigma}
    \right),
\end{equation}
where $b^2=a_{11}+a_{12}^2/a_{11}$.
The coefficients are shown in \autoref{table:2}.
The table illustrates that the last term in \autoref{eq:4} is indeed very small,
and the single-determinant with doubly-occupied extended orbital is an adequate
description.
The breathing lowers the energy cost of double occupancy and renormalizes the
effective $U_{\rm eff}$.

\begin{table}[t]
  \caption{Coefficients of the wave function in \autoref{eq:4} for the isolated
    atom.\label{table:2}}
\begin{tabular}{ccc}
\hline
\hline
 $a_{11}$ &  $a_{12}$  &  $a_{22}-a_{12}^2/a_{11}$ \\
\hline
 0.91  & 0.29  & -0.01  \\
\hline
\hline
\end{tabular}
\end{table}

\subsubsection{Dimer}\label{sec:dimer}

\begin{figure}[!h]
  \includegraphics[width=.99\linewidth]{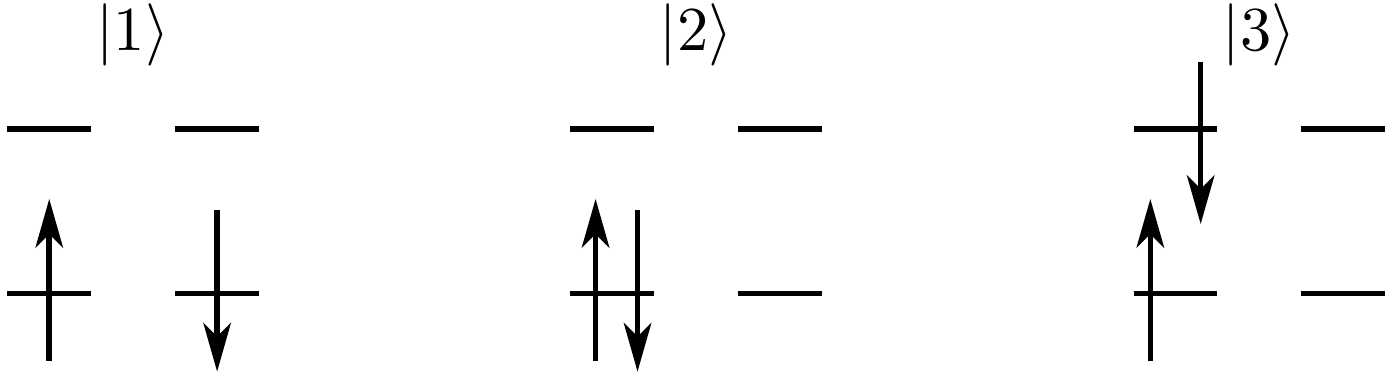}
\caption{Schematic representation of states in Eq.~(\ref{eq:5}) for the
dimer model.\label{fig:states}}
\end{figure}

We now turn to the full Cu$_2$ model.
\autoref{table:3} below shows the singlet-triplet splitting.
It illustrates how the inclusion of the integral $K$ strongly increases the
splitting, due to breathing effects.
To understand the results better, we consider a simpler model within only three
determinants for the singlet state.
\begin{eqnarray}\label{eq:5}
  && |1\rangle={1\over \sqrt{2}}\left(
    c^{\dagger}_{11\uparrow}c^{\dagger}_{21\downarrow}+
    c^{\dagger}_{21\uparrow}c^{\dagger}_{11\downarrow}
  \right)|{\rm vac}\rangle \nonumber \\
  && |2\rangle={1\over \sqrt{2}}\left(
    c^{\dagger}_{11\uparrow}c^{\dagger}_{11\downarrow}+
    c^{\dagger}_{21\uparrow}c^{\dagger}_{21\downarrow}
  \right)|{\rm vac}\rangle \\
  && |3\rangle={1\over 2}\left(
    c^{\dagger}_{11\uparrow}c^{\dagger}_{12\downarrow}+
    c^{\dagger}_{12\uparrow}c^{\dagger}_{11\downarrow}+
    c^{\dagger}_{21\uparrow}c^{\dagger}_{22\downarrow}+
    c^{\dagger}_{22\uparrow}c^{\dagger}_{21\downarrow}
  \right)|{\rm vac}\rangle \nonumber,
\end{eqnarray}
where $|{\rm vac}\rangle$ is the vacuum state with no electrons.
These basis states are shown schematically in \autoref{fig:states}.
State $|1\rangle$ corresponds to $d^9$-$d^9$ state mentioned in the main text,
while $|2\rangle$ and $|3\rangle$ resemble $d^8$-$d^{10}$ state in the main
state without and with 4$d$ occupation respectively.
Hamiltonian \eqref{eq:2} within the basis given by \autoref{eq:5} reads
\begin{equation}\label{eq:6}
  H=\left ( \begin{array}{ccc}
      2\varepsilon_1 & 2t_{11} & \sqrt{2}t_{12} \\
      2t_{11} & 2\varepsilon_1+U_{11}  & \sqrt{2}K_1 \\
      \sqrt{2}t_{12} & \sqrt{2}K_1 &\varepsilon_1+\varepsilon_2+U_{12}
\end{array} \right )
\end{equation}
Diagonalizing this matrix, we obtain the second column of in \autoref{table:3}.
These results agree rather well with the full calculation for the model in
\autoref{eq:2}, although the basis set in \autoref{eq:5} is incomplete.
The splitting is smaller because the higher states have been neglected.

\begin{table}[!h]
  \caption{Triplet-singlet splitting without ($K_1\!=\!0$) and with breathing.
    All values in eV.
    \label{table:3}}
\begin{tabular}{ccc}
\hline
\hline
$K_1$ &
       exact, \autoref{eq:2} & \autoref{eq:6}\\
\hline
0     &         0.077      &    0.076    \\
-8    &         0.176      &    0.167     \\
\hline
\hline
\end{tabular}
\end{table}
\newpage

We can now use L\"owdin folding, focusing on the upper $2 \times 2$ corner of
$(z-H)^{-1}$
\begin{equation}\label{eq:6a}
  (z-H)^{-1}\!=\!\left (\!
    \begin{array}{cc}
      z- 2\varepsilon_1- 2t_{12}^2/ \Delta E  & 2t_{11}-2t_{12}K_1/\Delta E \\
      2t_{11} -2t_{12}K_1/\Delta E & U_{11}-2K_1^2/\Delta E
\end{array}\!\right )^{-1}
\end{equation}
where $\Delta E=\varepsilon_2-\varepsilon_1+U_{12}$ and we have introduced the
approximation $z\approx 2 \varepsilon_1$ at some places.
The matrix in \autoref{eq:6a} shows rather clearly that there is an interference
between breathing and hopping from the 3$d$ orbital on one site to the 4$d$
orbital on the other site.
The effective value of $U$ has now been reduced
\begin{equation}\label{eq:6c}
  U_{11}  \to U_{11}^{\rm eff}\equiv U_{11}-2{K_1^2 \over \Delta E}
\end{equation}
and the effective hopping has been increased
\begin{equation}\label{eq:6d}
t_{11} \to t_{11}^{\rm eff}\equiv t_{11}-2{t_{12}K_1\over \Delta E},
\end{equation}
since $K_1<0$ and $t_{11}$ and $t_{12}$ have the same sign.
For the triplet case the basis state $|2\rangle$ does not exist, and these
renormalization effects are not present.
The singlet-triplet splitting is then
\begin{equation}\label{eq:6b}
  E_T-E_S\approx {4 \left(t_{11}-t_{12}K_1/\Delta E \right)^2 \over
    U_{11}-2K_1^2 / \Delta E}\equiv
  4 {\left(t_{11}^{\rm eff}\right)^2 \over U_{11}^{\rm eff}}.
\end{equation}
This illustrates the importance of the renormalization of $U_{11}$ and $t_{11}$.

\newpage
\section{Hopping for \element{Cu}$_2$O model}\label{sec:cu2O}

\begin{figure}[t]
  \includegraphics[width=.98\linewidth]{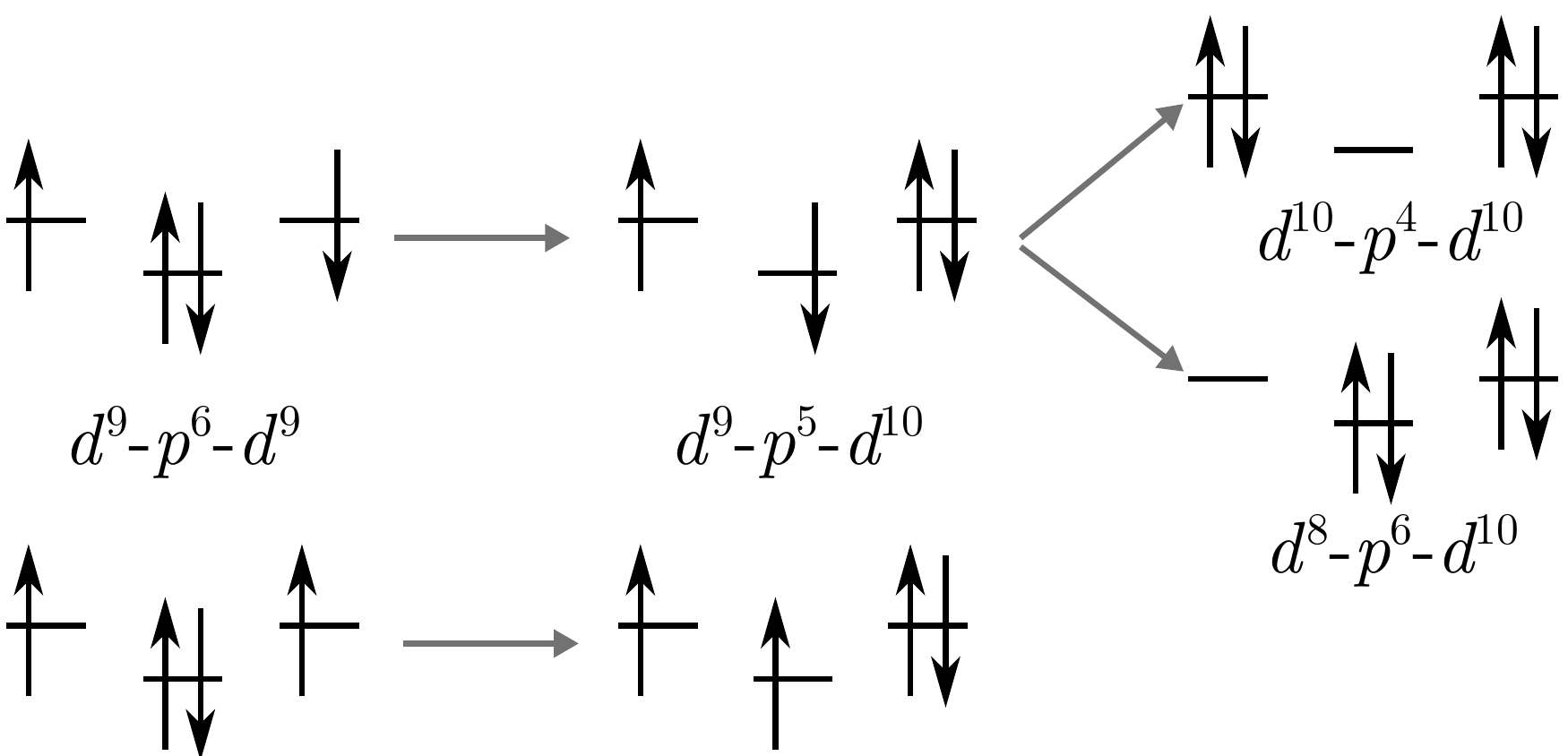}
\caption{
  Schematic representation of the Cu$_2$O model with Cu 3$d$ and O 2$p$ levels.
  The O atom is the bridging atom.
  The upper (lower) part of the figure illustrate available configurations for
  the singlet (triplet) state.
  \label{fig:superexchange}}
\end{figure}

\autoref{fig:superexchange} illustrate hopping possibilities for the singlet and
triplet state in a Cu$_2$O model with nondegenerate levels on Cu (3$d$) and O
(2$p$).
For the triplet case the hopping possibilities are severely limited.
In this model $J$ is given by
\begin{equation}\label{eq:11}
  J={4t_{\rm eff}^2\over \left(U_{\rm eff}+\Delta \right)^2}\left(
    {t^2\over U_{\rm eff}}+{t^2_{\rm eff}\over U_{\rm eff}+\Delta}\right),
\end{equation}
where $U$ is the $3d-3d$ Coulomb integral, $t$ is the hopping from Cu 3$d$ to O
2$p$ and $\Delta$ is the energy difference between the Cu 3$d$ orbital and the O
2$p$ orbital.
In a model with 4$d$ orbitals on the Cu atoms, $t$ is renormalized to
$t_{\rm eff}$ and $U$ to $U_{\rm eff}$.
\newpage

\end{document}